\documentclass[twoside]{article}  
\usepackage{amsfonts}
\usepackage{amsbsy}
\usepackage{amsmath}
\usepackage{algorithmic}
\usepackage{algorithm}
\usepackage[pdftex]{graphicx}
\usepackage{subfig}
\newcommand{\ruimte}{\par\vspace{1ex}\noindent}

\newcommand{\beq}{\begin{equation}}
\newcommand{\eeq}{\end{equation}}
\newcommand{\bmat}{\left[ \begin{array}}
\newcommand{\emat}{\end{array} \right]}

\title{\LARGE \bf
An algebraic approach to source coding with side information using list decoding}

\author{Mortuza Ali and Margreta Kuijper\footnote{M.\ Ali and M.\ Kuijper are with the Department of Electrical and Electronic Engineering, University of Melbourne, VIC 3010, Australia {\tt\small mortuza94@gmail.com; mkuijper@unimelb.edu.au}}%
\thanks{This work was presented in part in ISIT 2010. \newline This work was supported by the Australian Research Council(ARC).}}

\begin{document}
 
\maketitle
\begin{abstract}
Existing literature on source coding with side information (SCSI) mostly uses the state-of-the-art channel codes namely LDPC codes, turbo codes, and their variants and assume classical unique decoding. In this paper, we present an algebraic approach to SCSI based on the list decoding of the underlying channel codes. We show that the theoretical limit of SCSI can be achieved in the proposed list decoding based framework when the correlation between the source and side information is $q$-ary symmetric. We argue that, as opposed to channel coding, the correct sequence from the list produced by the list decoder can effectively be recovered in case of SCSI with a few CRC symbols. The CRC symbols, which allow the decoder to identify the correct sequence, incur negligible overhead for large block lengths. More importantly, these CRC symbols are not subject to noise since we are dealing with a virtual noisy channel rather than a real noisy channel. Finally, we present a guideline for designing constructive SCSI schemes for non-binary and binary sources using Reed Solomon codes and BCH codes, respectively. This guideline allows us to design a SCSI scheme for any arbitrary $q$-ary symmetric correlation without resorting to simulation.
\end{abstract}

\section{Introduction}
Recently, in the context of sensor networks and mobile multimedia applications~\cite{RamMagaDSC, LiveMagaDSC, GirodProcDVC}, distributed source coding has gained significant attention from the research community. Distributed source coding (DSC)~\cite{bookDSC} refers to the compression of correlated sources that are not co-located. Thus in a DSC setting, the encoding of correlated sources is performed independently while decoding is done jointly. The information theoretic limits for independent encoding of correlated sources have been established by Slepian and Wolf in~\cite{SlepianWolf}. According to the Slepian-Wolf theorem, independent encoding of correlated sources with joint decoding can be as efficient as joint encoding and decoding. More specifically, in compressing two correlated sources $X$ and $Y$, the rates achievable with independent encoding but joint decoding are bounded by $R_X \geq H(X|Y), R_Y \geq H(Y|X)$, and $R_X + R_Y \geq H(X,Y)$. In this paper, we focus on the asymmetric approach where $Y$ is encoded at a rate $H(Y)$ in the conventional way and $X$ is encoded at a rate $H(X|Y)$ assuming that $Y$ is available at the decoder. This asymmetric approach is known as {\em source coding with (decoder only) side information}~(SCSI) in the literature.

\ruimte
The essential technique for source coding with decoder only side information is {\em binning}. Consider the encoding of a source $X$ in the presence of the side information $Y$ available only at the decoder. Let $\mathbb{X}$ and $\mathbb{Y}$ be the alphabets of $X$ and $Y$ respectively. For large enough $n$, with high probability, a source sequence  $\mathbf{x} \in \mathbb{X}^n$ belongs to a set of approximately $2^{nH(X|Y)}$ sequences that are jointly typical with the side information sequence $\mathbf{y} \in \mathbb{Y}^n$. Thus if $\mathbf{y}$ were available both at the encoder and decoder, the outcomes from the source $X$ could be encoded using approximately $H(X|Y)$ bits on average with a very small probability of error. In this case, both the encoder and decoder could construct the same set of jointly typical sequences and use the same indexing. Then encoding of a source sequence $\mathbf{x}$ with its index would lead to correct decoding. However, even if $Y$ is not available at the encoder, it is possible to achieve the same rate of $H(X|Y)$ using the technique of binning~\cite{CovThoEleInfo5}. The idea is to randomly assign each of the source sequences in $\mathbb{X}^n$ to one of the $2^{nR}$ bins, where $R>H(X|Y)$. Given a source sequence $\mathbf{x}$, the encoding operation is to transmit the index of the bin to which $\mathbf{x}$ belongs. The decoding operation is to choose the sequence $\hat{\mathbf{x}}$ from the indexed bin which is jointly typical with the side information sequence $\mathbf{y}$. Since, for large enough $n$, almost all the sequences that are jointly typical with a given $\mathbf{y}$ will belong to different bins, $\hat{\mathbf{x}}$ will be equal to $\mathbf{x}$ with high probability.

\ruimte
Although the Slepian-Wolf theorem states the fundamental limit on achievable compression, a practical coding scheme to achieve this limit does not follow immediately. However, it follows from the above binning scheme that a practical binning algorithm needs to partition the source data space into a minimum number of bins while ensuing correct decoding. In other words, it should put as many source sequences as possible in a bin while ensuring that for any typical side information sequence each of the bins contains only one jointly typical source sequence. Thus each of the bins can play the role of a good channel code. This connection between binning and channel codes was first indicated in~\cite{RecResultWyner} while interpreting the Slepian-Wolf coding. 

\ruimte
Due to this close connection between binning and channel coding, most of the SCSI schemes proposed in the literature are based on the state-of-the-art channel codes namely LDPC codes, turbo codes, and their variants. Although their performances have been reported to be close to the theoretical limits, there are several obstacles in using them in practice. Firstly, the LDPC code based schemes~\cite{LiverisLDPC,GirodRateAdaLDPC,GirodRateAdaElsevier} and turbo code based schemes~\cite{TurboFriasDCC,TurboFriasLett,TurboMitranGCOM,TurboMitranSym,TurboGirodDCC} mostly focus on binary sources. Although some compression schemes for non-binary sources have been proposed~\cite{TurboFriasDCC,TurboGirodDCC,nonbinTurboJour}, they mainly map the non-binary symbols to fixed-length bit sequences and then encode the bit sequences using binary coding constructs. However, this requires decomposing the symbol level correlation into bit level correlation. 

\ruimte
Secondly, the near optimal performance of LDPC and turbo codes, as channel codes, can be attributed to their efficient soft-decision decoding algorithms. In soft-decision decoding, the soft output from a channel is directly fed into the decoder without demodulating to discrete values. Since demodulation of the received signal to hard values results in irretrievable loss of information, the soft-decision decoding results in approximately 2dB gain over its hard-decision decoding counterparts. However, it follows from the binning framework that the virtual channel essentially models the correlation between the source and side-information and thus dictates the choice of the channel code that induces an efficient binning. Clearly, the processes of modulation and demodulation are not involved in the SCSI. Therefore, in LDPC and turbo code based SCSI schemes, the soft-decision coding gain is not readily available.

\ruimte
Finally, there is always a gap between the compression rate that can be achieved with a specific channel code and the conditional entropy for which it can yield near lossless compression. For example, the turbo code based scheme in~\cite{TurboMitranGCOM} has a compression rate of $0.67$ but can achieve near lossless compression only when the true conditional entropy is $H(X|Y)=0.49$. Similarly the LDPC code based scheme in~\cite{LiverisLDPC}, that achieves near lossless compression when $H(X|Y)=0.20$, has an effective compression rate of $0.25$. However, for these schemes there is no algebraic approach to choose a code of a particular rate, given a conditional entropy, that can ensure near lossless recovery. Instead simulation is used to design a code, given the conditional entropy, due to the lack of algebraic construction of LDPC and turbo codes. 


\ruimte
In this paper, we present an algebraic approach to SCSI based on list decoding of the corresponding channel code that induces the partitioning of the source space. We demonstrate that for $q$-ary symmetric correlation between the source and side information, the list decoding approach can achieve optimal compression.
Although list decoding yields a list of sequences, as opposed to the classical unique decoding, we show that in case of SCSI, the correct sequence can be effectively extracted from the list at a negligible increase in the optimal compression rate. One of the advantages of using list decoding is that it can reduce the compression rate significantly as compared to its classical unique decoding counterpart. Besides, this algebraic approach allows us to design a SCSI scheme for any arbitrary $q$-ary symmetric correlation without resorting to simulation. 

\ruimte
Although there are extensive research activities in the area of list decoding and the same holds for SCSI, only a couple of works~\cite{CCSC-ISIT} can be found in the literature on the usage of list decoding for source coding problems. The use of list decoding has recently been proposed by Draper and Martinian in~\cite{CCSC-ISIT} for {\em compound conditional source coding} (CCSC).
As a variant of Slepian-Wolf coding, CCSC restricts the possible side information sequences to only a small subset of the jointly typical set. More importantly, the CCSC protocol presented in ~\cite{CCSC-ISIT} is non-constructive - a practical coding scheme does not follow immediately from it. In contrast, the scheme proposed in this paper addresses the broader problem of SCSI without putting any constraint on the size of the set of possible side information sequences. Moreover, the proposed approach clearly articulates the guideline to choose an appropriate channel code to design a practical SCSI scheme. 

\ruimte
After the publication of our preliminary work~\cite{aliK10}, that first proposed the usage of list decoding for SCSI, Li and Ramamoorthy~\cite{LiRamamoorthy11} suggested using RS codes in the context of SCSI with feedback. They proposed using soft-decision list decoding for better performance and used simulation for constructive code design. However, we demonstrate that computationally complex soft-decision decoding is not justified in the SCSI setting. We also show that SCSI schemes can be constructed algebraically using RS codes and BCH codes by making maximum use of their known list decoding bounds. Not surprisingly, later in this paper, we demonstrate that our algebraic approach with hard-decision list decoding achieves the same compression as that achieved by the approach in~\cite{LiRamamoorthy11}.


\ruimte
The organization of the rest of the paper is as follows. In Section~\ref{sec:syndrome}, we briefly review the technique of {\em syndrome source coding} which provides a framework for designing SCSI schemes using linear block codes. In Section~\ref{sec:proposed-scheme} we present the notion of list decoding and describe how this can be used to design a SCSI scheme based on the technique of syndrome source coding. The usage of Reed-Solomon codes and BCH codes in the design of practical constructive codes for SCSI is detailed in Section~\ref{sec:constructive}. Finally we conclude the paper in Section~\ref{sec:con}.

\section{Syndrome Source Coding for $q$-ary Symmetric Correlation}
\label{sec:syndrome}
The most challenging problem in the design of a practical binning scheme is the systematic construction of bins with rich algebraic structures so that the bin indexing and typical set decoding can be performed with reasonable complexity. In this regard, there is a close connection between binning and channel coding. A linear block code induces a partitioning of the source space into cosets that can be indexed by their respective syndromes~\cite{SloaneTheoErroCode}. If the cosets of a linear code are such that each of them with high probability contains only one sequence from the typical set then the cosets effectively act as bins. In this case, the index of the bin in which a source sequence belongs can be computed as the syndrome of the sequence.
\begin{figure}[!tb]
\centering
\includegraphics[width=9.5cm]{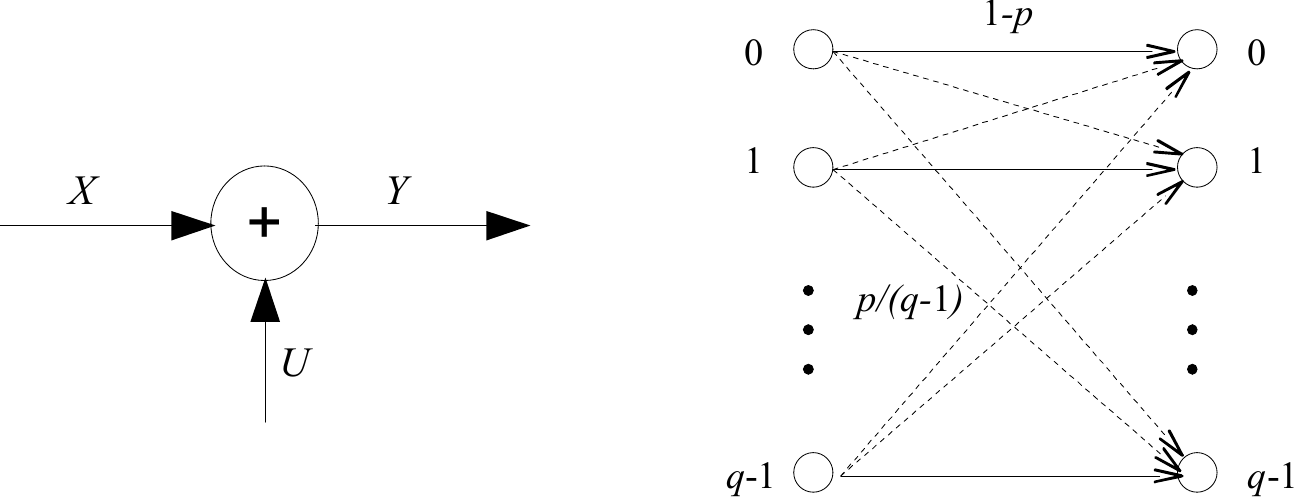}
\caption{The source $X$ and side information $Y$ is correlated as $Y=X+U$ where $U$ is an IID $q$-ary source such that $\text{Pr}[U=0]=1-p$ and $\text{Pr}[U=u]=p/(q-1)$, for any $u \in GF(q)-\{0\}$. Thus this correlation between $X$ and $Y$ is modelled via a $q$-ary symmetric channel with error probability $p$.}
\label{fig:noise-model}
\end{figure}

\ruimte
Consider encoding a memoryless $q$-ary source $X$ with correlated side information $Y$ available only at the decoder such that
\begin{equation*}
Y = X + U,
\end{equation*}
where $U$ is an IID $q$-ary source such that 
\begin{equation*}
\text{Pr}[U=u] = \left \{
\begin{array}{ll}
	1-p & \text{if } u=0\\
	p/(1-q) & \text{if } u \in GF(q)-\{0\}
\end{array} \right.
\end{equation*}
If $Y$ were present at the encoder as well, $X$ could be compressed at a rate of $H_q(X|Y) = H_q(U) = -p \log_q p - (1-p) \log_q (1-p) + p \log_q (q-1)$ symbols. According to the Slepian-Wolf theorem, $X$ can be compressed at the same rate $H_q(U)$ even if $Y$ is present only at the decoder. 

\ruimte
The above SCSI scenario can be modelled using a virtual $q$-ary symmetric channel ($qSC$) with error probability $p$, which for a $q$-ary symbol input, outputs the same input symbol with probability $1-p$, or any of the other $q-1$ symbols with probability $p/(q-1)$ (see Fig.~\ref{fig:noise-model}). This modelling of the correlation between the source and side information with a virtual channel allows us to use a channel code for the $qSC$ to design a SCSI scheme~\cite{ZamirNestedLattice} as described below.
The capacity $C_{qSC}$ of the $qSC$ and equivalently of the additive noise channel is
\[
C_{qSC} = 1 - H_q(U).
\]
According to the channel coding theorem~\cite{MackayInfoTheoInfAlg4}, there exists a $(n, k)$ linear block code $C$ over $GF(q)$ of rate $k/n = R > (C_{qSC} - \delta)$ such that the decoding error probability $P_e < \epsilon$ for any  $\epsilon > 0$ and $\delta > 0$. Here the error event $\mathbf{x} \neq \hat{\mathbf{x}}$ corresponds to the fact that when the actual noise vector is $\mathbf{u}$, the decoder decides the noise vector to be $\hat{\mathbf{u}}$ with $\mathbf{u} \neq \hat{\mathbf{u}}$. Therefore, $\textnormal{Pr}(\mathbf{x} \neq \hat{\mathbf{x}}) = \textnormal{Pr}(\mathbf{u} \neq \hat{\mathbf{u}})$. 

\ruimte
Now consider the following scheme of compression of $X$ with the side information $Y$ available only at the decoder based on this channel code. The encoding operation is to compute the $(n-k)$-symbol syndrome $\mathbf{s}_{\mathbf{x}}$ of the source sequence $\mathbf{x} \in \mathbb{X}^n$ as $\mathbf{s}_{\mathbf{x}} = \mathbf{H}\mathbf{x}^T$, where $\mathbf{H}$ is the parity check matrix of $C$. If $C_{\mathbf{s}_{\mathbf{x}}}$ denotes the coset corresponding to the syndrome $\mathbf{s}_\mathbf{x}$, then clearly $\mathbf{x} \in C_{{\mathbf{s}_\mathbf{x}}}$. The decoding operation is to find the sequence $\mathbf{\hat{x}}$ nearest (in Hamming distance) to the side information sequence $\mathbf{y} \in \mathbb{Y}^n$ in the coset $C_{\mathbf{s}_\mathbf{x}}$. Since $Y=X+U$, we can compute the syndrome of $\mathbf{u}$ as $\mathbf{s}_{\mathbf{u}} = \mathbf{H}\mathbf{y}^T - \mathbf{s}_{\mathbf{x}}$. Then the nearest neighbour decoding is equivalent to finding the minimum-weight noise vector $\hat{\mathbf{u}} \in C_{\mathbf{s}_\mathbf{u}}$ and decode $\mathbf{x}$ as $\hat{\mathbf{x}} = \mathbf{y} - \hat{\mathbf{u}}$. Thus the probability of error of the scheme is $\textnormal{Pr}(\mathbf{x} \neq \hat{\mathbf{x}}) = \textnormal{Pr}(\mathbf{u} \neq \hat{\mathbf{u}})$, same as the channel decoding error, which tends to zero as $n \rightarrow \infty$. This coding scheme, which can compress $X$ at a rate of $(n-k)/n< H_q(U) + \delta$ symbols with an arbitrarily small probability of error, is known as {\em syndrome source coding}~\cite{AnchetaSyndrome}. Clearly, if a channel code of rate $R=k/n$ is used for syndrome source coding, the achieved compression rate is $(n-k)/n=1-R$.

\section{Syndrome Source Coding Using List Decoding}
\label{sec:proposed-scheme} 
\begin{figure}[!tb]
\centering
\includegraphics[width=4.5cm]{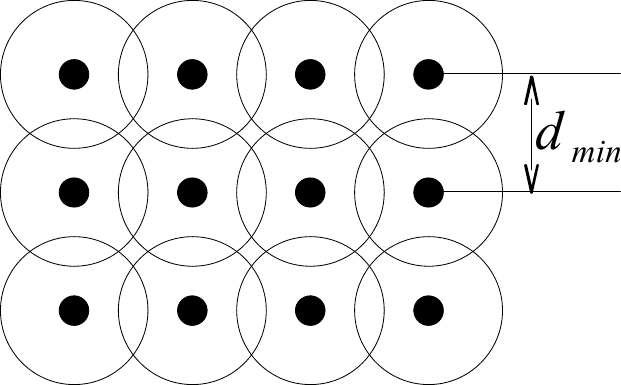}
\caption{Unambiguous bounded distance decoding (BDD) is possible only if $\tau < d_{\min}/2$}
\label{fig:beyond-dmin}
\end{figure}

Clearly, the underlying linear block code $C$ and its associated decoding algorithm impact the performance of a syndrome source coder. In the decoding of channel codes, the objective is to find the transmitted codeword $\mathbf{c} \in C$, given the received word $\mathbf{r} \in GF(q)^n$. The natural decoding approach is to find the codeword which has the maximum likelihood of being transmitted given that $\mathbf{r}$ has been received. This approach known as {\em maximum likelihood decoding}~(MLD) amounts to finding the codeword $\hat{\mathbf{c}}$ closest to $\mathbf{r}$ in an appropriate measure  of distance. However, MLD is known to be NP-complete in general~\cite{MDD-NP-complete}. In practice, {\em bounded distance decoding} (BDD), which has greatly reduced complexity, is preferred that ensures correct decoding only when the number of errors is upper bounded by some error correcting radius $\tau$. Obviously, an unambiguous BDD is possible only if $\tau < d_{\min}/2$ where $d_{\min}$ is the minimum distance of the code $C$~(see Fig.~\ref{fig:beyond-dmin}).

\ruimte
Let us see the implication of BDD for channel coding and in turn for syndrome source coding. For a $qSC$ with error probability $p$, the expected Hamming distance between the transmitted codeword $\mathbf{c}$ and the received word $\mathbf{r}$ is $E[d(\mathbf{c}, \mathbf{r})]=np$. Thus for unambiguous decoding, we need a code $C$ with $d_{\min}>2np$. This leads to the question of the largest rate possible with a code whose minimum distance is $2np$. This remains as one of the biggest open questions in combinatorial coding theory~\cite{binaryListGuru}. However, it is known that the rate achievable with a block code of minimum distance $2np$ is much less than $1-H_q(p)$. For example, the best known upper bound on the rate achievable with a binary code of block length $n$, having a minimum distance of $2np$, is less than $1 - H_2(2p)$. Note that that the rate becomes $0$ when $p > 0.25$.
This in turn implies that the compression rate achievable with syndrome source coding that relies on BDD is lower bounded by $H_2(2p)$, which is strictly larger than $H_2(p)$.

\ruimte
In the above scenario, the main constraint is the requirement of unique decoding which sets the decoding radius to $\lfloor (d_{\min}-1)/2 \rfloor$. One way to circumvent this limitation is to increase the decoding radius beyond $\lfloor (d_{\min}-1)/2 \rfloor$ and allow the decoder to output a list of codewords. This approach will be feasible as long as (i) the list contains only a small number of codewords and (ii) there is an effective way of extracting the correct codeword from the list. The method of decoding beyond $\lfloor (d_{\min}-1)/2 \rfloor$ is known as list decoding in the literature. Let $\mathbb{B}_q(\mathbf{r}, e)$ denote the Hamming sphere of radius $e$ around  a point $\mathbf{r}$ in the space $GF(q)^n$. A code $C$ over $GF(q)$ is said to be $(p, L)$ list-decodable if $|\mathbb{B}_q(\mathbf{r}, np) \cap C| \leq L$. List decoding is considered feasible as long as $L$ is a polynomial in~$n$.

\ruimte
To assess the feasibility of using list decoding for SCSI, let us look at the theoretical limits on list decoding. It has been shown in~\cite{algorithmic-ListDecode} that for any integer $L \geq 2$, there exists a family of binary linear  $(p, L)$ list-decodable channel codes of rate $R \geq 1 - H_2(p)- 1/L$. Thus allowing $L$ to grow, a rate arbitrarily close to the theoretical limit $1 - H_2(p)$ can be achieved. This in turns implies that the corresponding syndrome source coders have compression rate $\leq H_2(p) + 1/L$, which can be made arbitrarily close to the conditional entropy $H_2(p)$ by allowing $L$ to grow. For a non-binary alphabet, a similar reasoning can be applied. It has been shown in~\cite{Guru-nonbinary-linear} that for any alphabet size of $q \geq 2$, list size $L \geq 2$, and $p \in (0, 1-1/q)$, there exists a family of $(p, L)$ list-decodable $q$-ary linear channel codes of rate $R \geq 1 - H_q(p) - 1/L$. Consequently, the corresponding $q$-ary syndrome source coders have a compression rate $\leq H_q(p)+1/L$ which approaches $H_q(p)$ as the list size $L$ increases.

\subsection{A geometrical interpretation of syndrome source coding using list decoding} 
\begin{figure}[!tb]
\centering
\includegraphics[width=5.5cm]{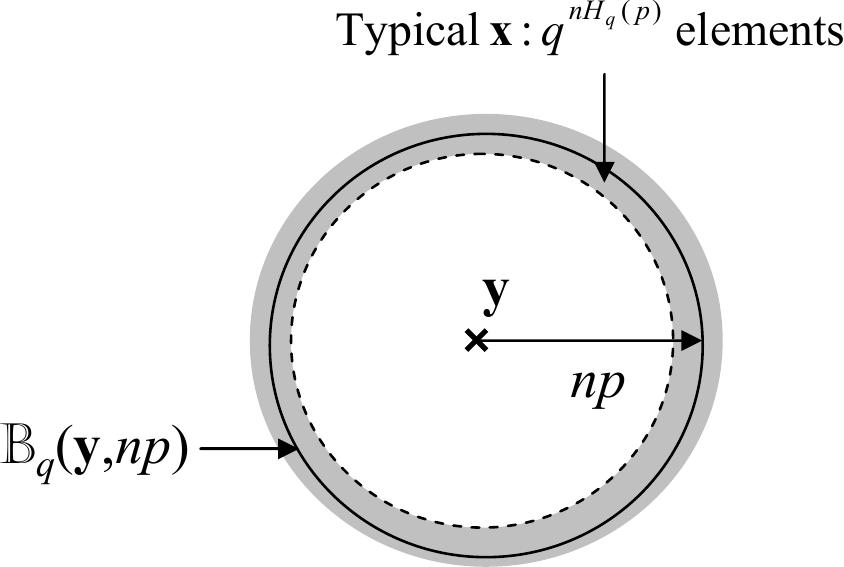}
\caption{For a given side information sequence $\mathbf{y} \in GF(q)^n$, the source sequence $\mathbf{x} \in GF(q)^n$, with high probability, will be within the thin shell on the surface of $\mathbb{B}_q(\mathbf{y}, np)$.  Total number of points in the shell is $|\mathbb{B}_q(\mathbf{y}, np)| \leq  q^{nH_q(p)}$.}
\label{fig:geom-inter}
\end{figure}

Consider the $q$-ary symmetric correlation between the source and side information with error probability~$p$. According to the law of large numbers, for large enough $n$, given a side information sequence $\mathbf{y} \in GF(q)^n$, the source sequence $\mathbf{x} \in GF(q)^n$ with high probability will be within a thin shell on the surface of $\mathbb{B}_q(\mathbf{y}, np)$. In fact, the thin shell corresponds to the set of sequences that are jointly typical with $\mathbf{y}$. Now the total number of points in the shell is approximately equal to $|\mathbb{B}_q(\mathbf{y}, np)|$ since for large $n$ almost all the point in $\mathbb{B}_q(\mathbf{y}, np)$ will be in the thin shell (see 
Fig.~\ref{fig:geom-inter}). It is known~\cite{algorithmic-ListDecode} that the number of points contained in $\mathbb{B}_q(\mathbf{y}, np)$ is bounded by
\begin{equation}
\label{eqn:spherebound}
|\mathbb{B}_q(\mathbf{y}, np)| \leq q^{nH_q(p)} .
\end{equation}
Now consider syndrome encoding of $\mathbf{x}$ using an $(n, k)$ channel code $C$ over $GF(q)$. Clearly $C$ induces a partitioning of the source data space $GF(q)^n$ into $q^{n-k}$ cosets. Since the points in the shell are uniformly distributed over $GF(q)^n$, for any syndrome $\mathbf{s}$, the number of points in $\mathbb{B}_q(\mathbf{y}, np) \cap C_{\mathbf{s}}$ is approximately $q^{nH_q(p)}/q^{n-k}$. From the list decoding point of view, this syndrome source coding is feasible if $|\mathbb{B}_q(\mathbf{y}, np) \cap C_{\mathbf{s}}|$ is small, which holds only if $R=k/n \leq 1 - H_q(p)$.

\subsection{Extracting the correct sequence from the list} In syndrome source coding based on list decoding, the decoder essentially produces a list of sequences. Thus after list decoding we need to choose the correct sequence from the list. To aid the decoder in finding the correct sequence from the list, we propose to transmit a few CRC symbols along with the syndrome. A CRC-$\rho$ code is defined by a generator polynomial $g(\xi)$ of degree $\rho$ that assigns $\rho$-symbol parity to a sequence. CRC codes are widely used in practice for error detection. In the setting of list-decoding based syndrome source coding, the use of a CRC code is expected to be effective for at least two reasons. Firstly, while in the context of channel coding, the CRC symbols are also subject to channel noise, this is not the case for syndrome source coding. In SCSI we are dealing with a virtual channel, modelling the correlation, instead of a real noisy channel. Thus in syndrome source coding, we can assume that these CRC symbols along with the syndrome will be available to the decoder without error. Secondly, since the list size $L$ is small (polynomial in $n$), only a few parity symbols are sufficient to correctly identify the desired sequence with high probability. Clearly a CRC-$\rho$ code identifies the correct sequence from a list of size~$L$ with high probability if $\rho \geq \log_q L$. Thus a syndrome source coder based on list decoding that uses a CRC code to extract the correct sequence has a compression rate of $H_q(p)+1/L + \log_q L/n$. Since $L$ is a polynomial in $n$, the compression rate $H_q(p)+1/L + \log_q L/n$ approaches to $H_q(p)$ as $n$ grows.


\subsection{The proposed framework}
Given a linear $(p, L)$ list-decodable code $C$ over $GF(q)$ of rate $>1 - H_q(p)$, let us articulate the encoding and decoding operations involved in the syndrome coding of $\mathbf{x}$ in the presence of side information $\mathbf{y}$ only at the decoder. Let $\mathbf{H}$ be the parity check matrix of the code $C$.

\ruimte
{\bf Encoding: } The syndrome of the source sequence $\mathbf{x}$ is computed as $\mathbf{s}_\mathbf{x} = \mathbf{H}\mathbf{x}^T$. Let $g(\xi)$ be the generator polynomial of a CRC-$\rho$ code and $x(\xi)$ be the polynomial of degree at most $n-1$ that corresponds to the sequence $\mathbf{x}$. Then the $\rho$-symbol CRC of $\mathbf{x}$ corresponds to the polynomial $h(\xi) = x(\xi) \mod g(\xi)$ of degree at most $\rho-1$.

\ruimte
{\bf Decoding:} First we need to list decode $C_{\mathbf{s}_\mathbf{x}}$ considering the side information $\mathbf{y}$ as the received word. There can be two approaches. In the first approach, we compute the syndrome of the noise vector $\mathbf{u}$ as $\mathbf{s}_\mathbf{u} = \mathbf{H}\mathbf{y}^T - \mathbf{s}_{\mathbf{x}}$. Then using a list decoding algorithm we determine the first $L$ minimum-weight sequences $\{\mathbf{u}_1, \mathbf{u}_2, \cdots, \mathbf{u}_L\}$ from the coset $C_{\mathbf{s}_\mathbf{u}}$. Subtracting each of the elements $\mathbf{u}_i, 1 \leq i \leq L$, from $\mathbf{y}$ we get the list $\mathcal{L}_{\mathbf{s}_\mathbf{x}}$. The list $\mathcal{L}_{\mathbf{s}_\mathbf{x}}$ essentially consists of the sequences from $C_{\mathbf{s}_\mathbf{x}}$ that are at a Hamming distance $\leq np$ from $\mathbf{y}$. In the second approach we first determine a sequence $\mathbf{a} \in C_{\mathbf{s}_\mathbf{x}}$. Then given $\mathbf{a} \in C_{\mathbf{s}_\mathbf{x}}$, the list decoding algorithm for $C$ can be used to construct the list $\mathcal{L}_{\mathbf{s}_\mathbf{x}}$ as follows. Compute $\mathbf{y}^\prime = \mathbf{y} - \mathbf{a}$. Using the list decoding algorithm for $C$, determine the list $\mathcal{L}$ consisting of those codewords of $C$ that are within the Hamming sphere of radius $np$ around $\mathbf{y}^\prime$. Then adding $\mathbf{a}$ to each of the codewords in $\mathcal{L}$, we get the list $\mathcal{L}_{\mathbf{s}_\mathbf{x}}$ consisting of the sequences from $C_{\mathbf{s}_\mathbf{x}}$ that are at a Hamming distance $\leq np$ from $\mathbf{y}$. 

\ruimte
{\em Remarks:} Between the two approaches, the first approach is preferred if the list decoding algorithm for $C$ starts from the syndrome of the received word and compute the first $L$ minimum weight sequences corresponding to that syndrome. Wu's list decoding algorithms for Reed-Solomon codes and BCH codes~\cite{WuListRS-BCH} are examples of such algorithms.  On the other hand, if the list decoding algorithm for $C$ does not start from the syndrome, the second approach becomes useful\footnote{In fact, the second approach can be adopted with any list decoding algorithm for $C$.}. For example, the Guruswami-Sudan list decoding algorithm~\cite{sudanGuruListRS} for RS codes does not rely on syndrome computation. However, the problem of finding a sequence $\mathbf{a} \in C_{\mathbf{s}_\mathbf{x}}$ amounts to solving the system of linear equations $\mathbf{H}\mathbf{a}^T = \mathbf{s}_\mathbf{x}$. Since there are more unknown variables than equations, there is at least one non-zero solution to it. Thus we can find an $\mathbf{a} \in C_{\mathbf{s}_\mathbf{x}}$ in polynomial time, for example, using Gaussian elimination.

\ruimte
Once we construct the list $\mathcal{L}_{\mathbf{s}_\mathbf{x}}$ using either of the two approaches, we pick the sequences $\hat{\mathbf{x}}$ from the list such that $\hat{x}(\xi) \mod g(\xi) = h(\xi)$. Decoding is considered successful if there is only one such sequence. An error event occurs if none of the sequences matches the CRC $h(\xi)$ or if there are more than one sequences having the same CRC $h(\xi)$. 

\begin{figure}[!tb]
\centering
\includegraphics[width=7.5cm]{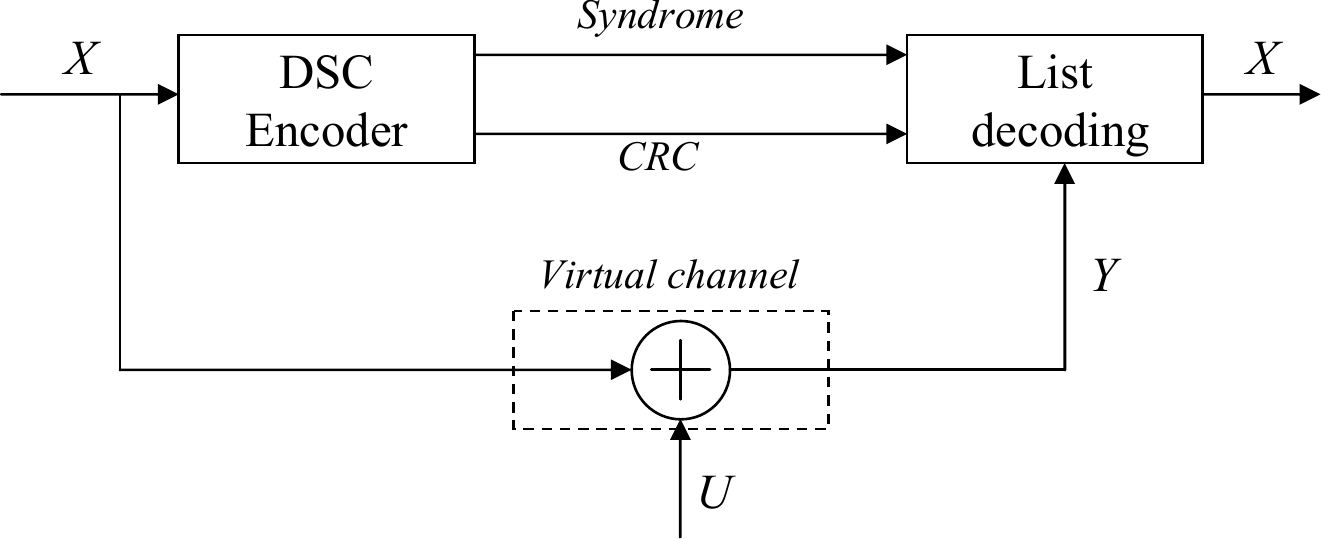}
\caption{The proposed framework. }
\label{fig:proposed}
\end{figure}

\ruimte
We summarize the proposed framework in Fig.~\ref{fig:proposed}. Here it is worth noting that, the channel only models the correlation between the source and side information and the syndrome and CRC are not subject to noisy transmission.

\section{Constructive Code Design}
\label{sec:constructive}
Although the encoding and decoding algorithms presented in the previous section are theoretically sound, there are at least two challenges while designing codes for real-world applications. Firstly, the scheme assumes that $d(\mathbf{x}, \mathbf{y})\approx np$ with high probability, which holds when $n \rightarrow \infty$. However, in the real world we have to operate with finite $n$. Secondly, the scheme also depends on the availability of a $(p, L)$ list-decodable code with an efficient encoder and decoder. To date, efficient list decoding algorithms are known for the families of Reed-Solomon (RS), Bose-Chaudhuri-Hocquenghem (BCH), and Reed-Muller (RM) codes. Associated with each of these codes $C$ is a known list decoding radius~$\tau$. Before presenting the main results on these families of codes and their potential in syndrome source coding, in the following we provide a general guideline that can be used to design a practical SCSI scheme for a given $q$-ary symmetric correlation specified by the error probability~$p$.

\ruimte
{\em Block length $n$:}  For better performance it is desirable to have $n$ as large as possible. However, for large $n$ the computational complexity may become impractical. While RS codes in practical applications are mostly of length $n=256$ (due to the byte oriented world), it is feasible to go up to $n=1024$ with binary BCH and RM codes.

\ruimte
{\em List decoding radius $\tau$:} In theory, for $n$ approaching $\infty$, a list decoding radius of $\tau<np+\delta$, where $\delta>0$, is sufficient. In practice, the required list decoding radius~$\tau$ depends on the desired error rate $\epsilon$. For fixed $n$ we need to have a list decoding radius of $\tau > T_\epsilon$, where $T_\epsilon$ is such that $\textnormal{Pr}(d(\mathbf{x}, \mathbf{y})>T_\epsilon)< \epsilon$. It can be shown that for large $n$ we need to choose $T_\epsilon$ slightly bigger than $np$. Let $e$ be a random variable representing $d(\mathbf{x}, \mathbf{y})$. Then clearly $e$ has a binomial distribution with mean $np$ and variance $\sqrt{(np(1-p))}$. For large $n$, with high probability $e$ will be in the vicinity of~$np$. For example, for $n=1000$, $p=0.2$, and $\epsilon=10^{-4}$, we have $T_\epsilon = 248$. If we want to decrease the error probability to $\epsilon = 10^{-5}$, we need to increase $T_\epsilon$ only by $8$ to $256$.

\ruimte
{\em Code rate $R$:} Since the compression rate achieved with syndrome source coding based on a channel code of rate $R$ is $1-R$, we need to pick a code of largest rate $R$ whose list decoding radius is just larger than $T_\epsilon$.

\ruimte
{\em CRC code generator $g(\xi)$:} There are a number of standard CRC codes, see~\cite{Wicker-CRC}. Since the list produced by the list decoder is guaranteed to be small, a few CRC symbols are enough to correctly extract the source sequence from the list with high probability. In practice, a 16-bit CRC is expected to be enough for $n=255$ and for sources over $GF(2^8)$ this would incur only $0.78\%$ overhead.

\ruimte
In the following we discuss the main list decoding results for the families of RS codes and BCH codes and their implication for syndrome source coding.

\subsection{Code design for non-binary sources using RS codes}
A $(n, k)$ RS code $C$ over $GF(q)$ can be defined with its generator matrix $\mathbf{G}$ and parity check matrix $\mathbf{H}$ as follows~\cite{Kotter-RS-GH-def} (here $1 \leq k \leq n = q-1$)
\begin{align*}
\{\mathbf{G}\}_{i, j}&= \alpha^{ij}, \quad i=0, \ldots, k-1, \ j = 0, \ldots, n-1 \\
\{\mathbf{H}\}_{i, j}&= \alpha^{(i+1)j}, \quad i=0, \ldots, n-k-1, \ j = 0, \ldots, n-1,
\end{align*}
where $\alpha$ is a primitive element of $GF(q)$. According to the generator matrix, the codeword corresponding to a message $\mathbf{u}$ can be computed as $\mathbf{c} = (u(\alpha^0), u(\alpha^1), \cdots, u(\alpha^{n-1}))$.
On the other hand, it follows from the parity check matrix that $\mathbf{c} \in C$ if and only if $c(\alpha^{i+1})=0$ for all $0 \leq i \leq n-k-1$. Further $d_{\min} = n-k+1$ since a $(n,k)$ RS code is {\em maximum distance separable} (MDS). 

\ruimte
A list decoding algorithm was first discovered for low rate RS codes by Sudan~\cite{sudanListDecodeRS} and later improved and extended for all rates by Guruswami and Sudan~\cite{sudanGuruListRS}. For a RS code of rate $R$, the Guruswami-Sudan algorithm can correct up to $n(1 - \sqrt{R})$ errors which is clearly beyond half of its minimum distance. The Guruswami-Sudan algorithm uses the polynomial representation corresponding to $\mathbf{G}$. Given a received word $\mathbf{r}$, the essential idea of the algorithm is to find the polynomials $u(\xi)$ of degree at most $k$ such that $u(\alpha^j)=r_i$ for at least $\tau$ values of $j\in [0, n-1]$. Recently, building upon the Guruswami-Sudan approach, Wu~\cite{WuListRS-BCH} has proposed an algorithm that can also achieve the same list decoding radius but with a reduced complexity. Moreover, the Wu algorithm, which relies on polynomial representation corresponding to $\mathbf{H}$, is well suited for syndrome source coding since its decoding algorithm, like the Berlekamp-Massey algorithm, starts with syndromes.

\ruimte
{\em Example:} Let us design a SCSI scheme for a source over $GF(q=2^8)$ where the correlation between the source and side-information is $q$-ary symmetric with error probability $p=0.3$. Given the desired error rate of $\epsilon = 10^{-4}$ and block length of $n=255$, the value of $T_\epsilon$ turns out to be $T_\epsilon = 105$. Thus we need a channel code of block length $n=255$ having list decoding radius $\tau > 105$. Using the fact that the RS code of rate $R$ has $\tau = n(1 - \sqrt{R})$, we find that the $(255, 88)$ is the desired RS code. Thus the compression rate achieved with this scheme is $1 - R = 0.6549$. When $16$ CRC bits are considered, the compression rate increases to $0.6627$. In contrast, with classical unique decoding we require a code to have $d_{\min}=2T_\epsilon +1=211$. It is the $(255, 45)$ RS code that has $d_{\min} = 211$. The use of this code for syndrome source coding with unique decoding can only achieve a compression rate of $0.8235$.

\subsection{Non-binary SCSI with feedback using RS codes}
Here we consider constructive code design for SCSI with feedback. In this setting, it is assumed that the decoder can request additional symbols via a feedback channel if the decoding is unsuccessful. A SCSI scheme can take advantage of the feedback channel as follows. It first uses a high rate channel code $C_1$ to induce the partition and transmits the syndrome $\mathbf{H}_1\mathbf{x}^T$. Upon receiving the syndrome, the decoder tries to decode. The decoder then informs the encoder whether the decoding is successful via the feedback channel. If the decoding is unsuccessful, the encoder chooses a lower rate code $C_2$ (increase the length of the syndrome) and transmits the syndrome $\mathbf{H}_2\mathbf{x}^T$. The encoder gradually increases the transmission rate by using channel codes of lower rates until the decoding is successful. 

\ruimte
In the setting of SCSI with feedback, the proposed list decoding approach based on RS codes is suitable for two reasons. Firstly, the CRC symbols transmitted along with the syndrome provide an effective way of detecting whether the decoding is successful. Clearly the CRC symbols need to be transmitted only once at the beginning of the process. Secondly, as pointed out in~\cite{LiRamamoorthy11}, the class of RS codes provides a natural rate adaptivity which stems from the nested structure of the parity check matrices of RS codes. Let $\mathbf{H}_1$ and $\mathbf{H}_2$ be the parity check matrices of RS code of rates $R_1=k_1/n$ and $R_2=k_2/n$ respectively, where $k_1>k_2$. Let the encoder first transmit the $n-k_1$ symbol syndrome $\mathbf{s}_{\mathbf{x}}^1=\mathbf{H}_1\mathbf{x}^T$ and suppose that the decoding is unsuccessful. Now if the encoder chooses the lower rate code for syndrome source coding, it needs to inform the decoder of the $n-k_2$ symbol syndrome $\mathbf{s}_{\mathbf{x}}^2=\mathbf{H}_2\mathbf{x}^T$. Since $\mathbf{H}_1$ is embedded in $\mathbf{H}_2$ as its first $n-k_1$ rows, the encoder can compute and transmit the $k_1-k_2$ additional syndromes as the inner product of $\mathbf{x}$ and the last $k_1-k_2$ rows of $\mathbf{H}_2$.

\ruimte
{\em Example:} Let us design a SCSI scheme for a source over $GF(q=2^8)$ where the correlation between the source and side-information can be modelled with a $q$-ary symmetric channel with error probability $p=0.34$. Since $q=2^8$, we choose the block length as $n=q-1=255$. Let the desired error rate is $\epsilon = 10^{-3}$. Let $e$ denotes the Hamming distance between the source sequence $\mathbf{x} \in GF(q)^{255}$ and the side-information sequence $\mathbf{y} \in GF(q)^{255}$. We first determine $l=3$ and $h=171$ such that $\text{Pr}[l \leq e \leq h]>1-\epsilon$. Let $C^i$ denotes the RS code of largest rate having list decoding radius $\tau \geq i$. Then clearly $C^l \supset C^{l+1} \supset \cdots \supset C^h$. We start the encoding with $C^3$ which is the $(255, 249)$ RS code. If the decoding is unsuccessful we choose $C^4$ which is the $(255, 247)$ RS code having list decoding radius $\tau \geq 4$. Thus we gradually increase the list decoding radius by decreasing the rate of the corresponding RS code when the decoding is unsuccessful. We declare a failure if the usage of $C^{171}$, which is the $(255,27)$ RS code, fails to provide correct decoding. Clearly the average compression rate achieved with this scheme is $\sum_{i=l}^{h}\text{Pr}[e=i](1-R^i)=0.5634$ symbols where $R^i$ denotes the rate of $C^i$.

\begin{figure}[!tb]
\centering
\includegraphics[width=7.0cm]{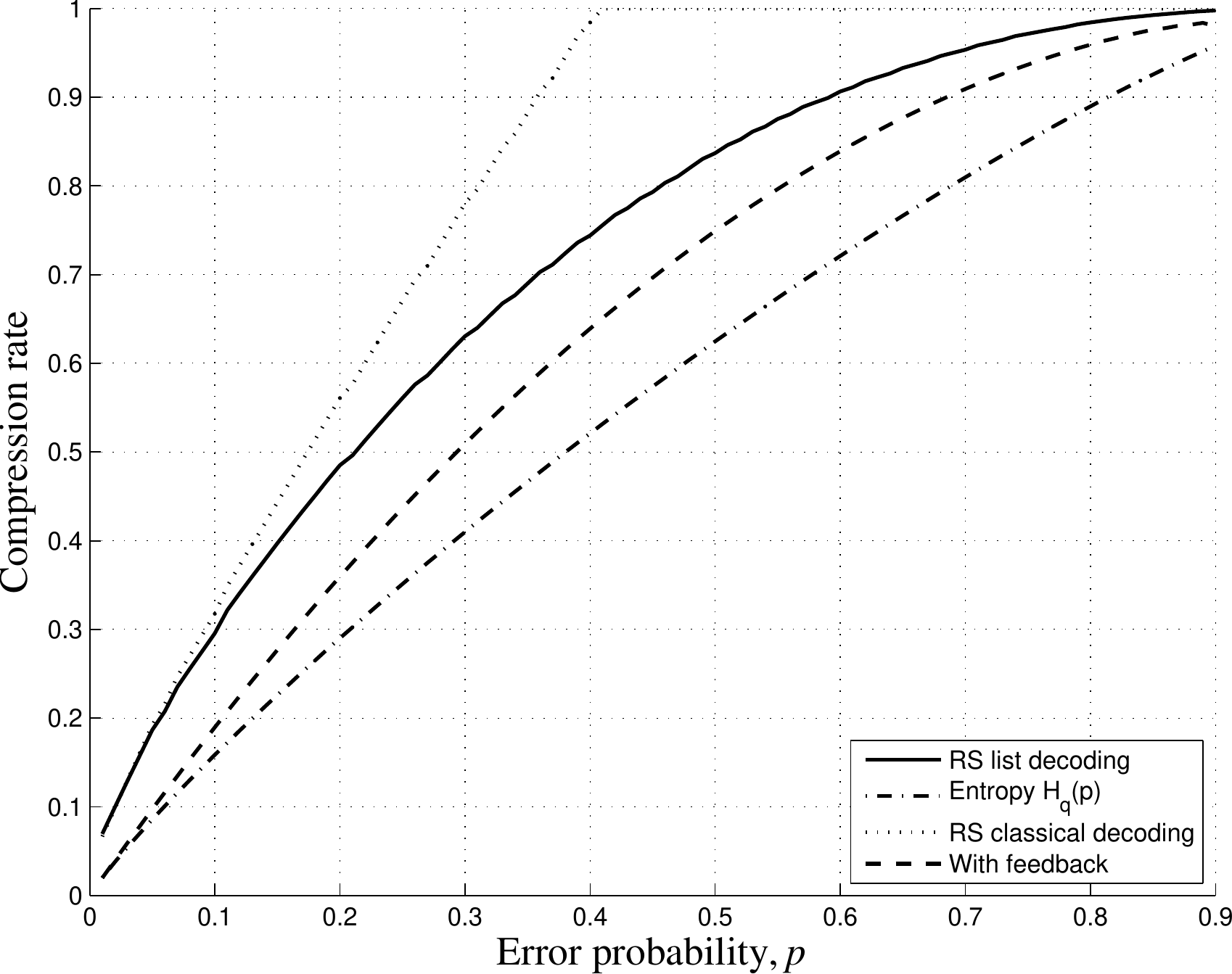}
\caption{Compression rate achievable with list decoding and unique decoding of RS codes under $q$-ary symmetric correlation specified by the error probability~$p$. Here $q=2^8$ and $\epsilon=10^{-3}$.}
\label{fig:SCSI-RS}
\end{figure}

\ruimte
In Fig.~\ref{fig:SCSI-RS}, we present the compression rates achievable with RS codes for sources over $GF(2^8)$. Here we have used the maximum block length $n=255$ and assumed $\epsilon=10^{-3}$ as the desired error rate. Clearly the SCSI scheme using list decoding of RS codes outperforms the scheme that relies on classical unique decoding. It is worth noting that for $p>0.4$, no compression can be achieved with unique decoding. For example, when $p=0.41$, the value of $T_\epsilon$ turns out to be $129$. With classical unique decoding to achieve a decoding error rate of $<10^{-3}$ we need a RS code of minimum distance $d_{\min}=259$ which does not exist for $n=255$. It also follows from the figure that the compression rate can be reduced further if a feedback channel is available to inform the encoder whether the decoding is successful.

\begin{figure}[!tb]
\centering
\includegraphics[width=7.0cm]{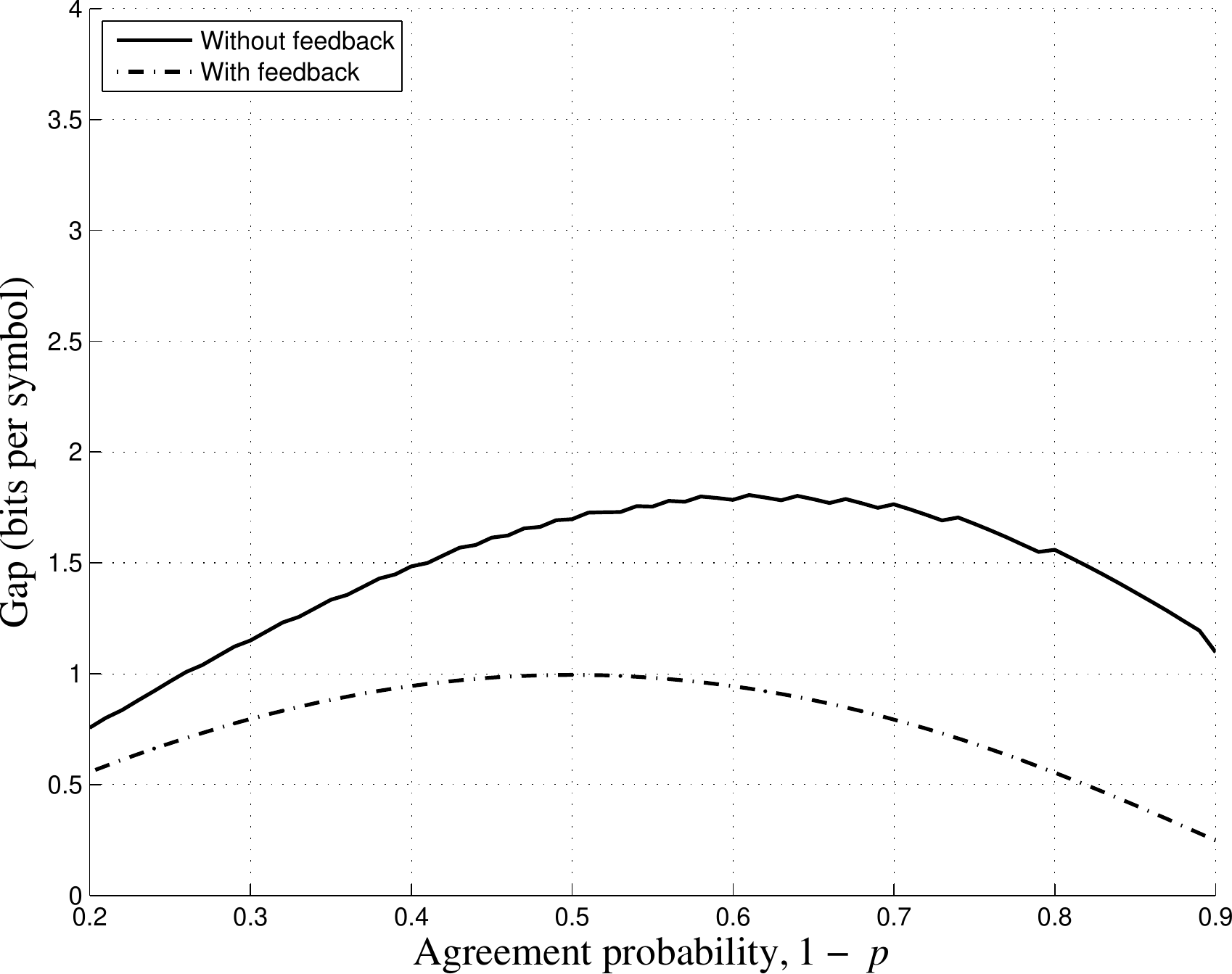}
\caption{The gap between $H(X|Y)$ and the transmission rate achievable with list decoding of RS codes for both the SCSI settings: classical (without feedback) and with feedback. Here $q=2^8$ and $\epsilon=10^{-3}$.}
\label{fig:gap-SCSI-RS}
\end{figure}

\ruimte
{\em Soft-decision vs hard-decision decoding:} In the SCSI setting the channel only models the correlation between the source and side-information and thus the processes of modulation and demodulation are not involved. Therefore, we argue that the computationally complex soft-decision decoding of RS codes, used for syndrome source coding, does not yield better performance over its hard-decision counterparts. To validate this we compare our results with that reported by Li and Ramamoorthy in~\cite{LiRamamoorthy11}. Li and Ramamoorthy, given a $q$-ary ($q=2^8$) correlation model specified by the error probability $p$, determined the required transmission rate to achieve an error rate of $\epsilon=10^{-3}$ via simulation. In simulation, Li and Ramamoorthy used soft-decision decoding of the corresponding RS codes. In~\cite{LiRamamoorthy11}, the simulation results have been reported as the gaps between the actual transmission rates and $H(X|Y)$ at different agreement probabilities ($1-p$). In Fig.~\ref{fig:gap-SCSI-RS}, we also present the gap between $H(X|Y)$ and the transmission rate calculated algebraically at different agreement probabilities $1-p$. Comparing Fig.~\ref{fig:gap-SCSI-RS} with that reported in~\cite{LiRamamoorthy11} we find that our algebraic approach yields the same rate as that achieved using simulation and soft-decision decoding. 

\subsection{Code design for binary sources using BCH codes}
Binary BCH codes can be interpreted as binary sub-codes of RS codes~\cite{SloaneTheoErroCode}, i.e., if $C_{RS}$ is an RS code over $GF(2^m)$, then $C_{RS} \cap GF(2)^n$ is a BCH code. This interpretation allows Wu's list decoding algorithm for RS codes to be used for the list decoding of BCH codes. In~\cite{WuListRS-BCH} Wu also presents an improved algorithm for list decoding of binary BCH codes that achieves a list decoding radius of $\tau = \frac{n}{2}(1 - \sqrt{1-2D})$, where $D$ is the designed relative distance of the BCH code~\cite{ECCCostello}.

\ruimte
{\em Example: } Consider designing a SCSI scheme for a $q$-ary symmetric correlation with error probability $p=0.2$ and $\epsilon=10^{-4}$. As binary BCH codes of length up to $1023$ can be implemented without any difficulty, we choose $n=1023$. For the given values of $n$, $p$, and $\epsilon$, we calculate $T_\epsilon = 254$. The $(1023, 56)$ BCH code with $D>0.3743$ has $\tau > 382$ and thus can achieve an error probability of $P_e<10^{-4}$ if used for syndrome source coding. The compression rate of this scheme is $0.9453$ which slightly increases to $0.9570$ when $12$ CRC bits are considered. Note that with unique decoding it would need a code of $d_{\min}>508$. The BCH code of designed distance $>508$ is the $(1023, 11)$ code which only achieves a compression rate of $0.9892$.

\begin{figure}[!tb]
\centering
\includegraphics[width=7.0cm]{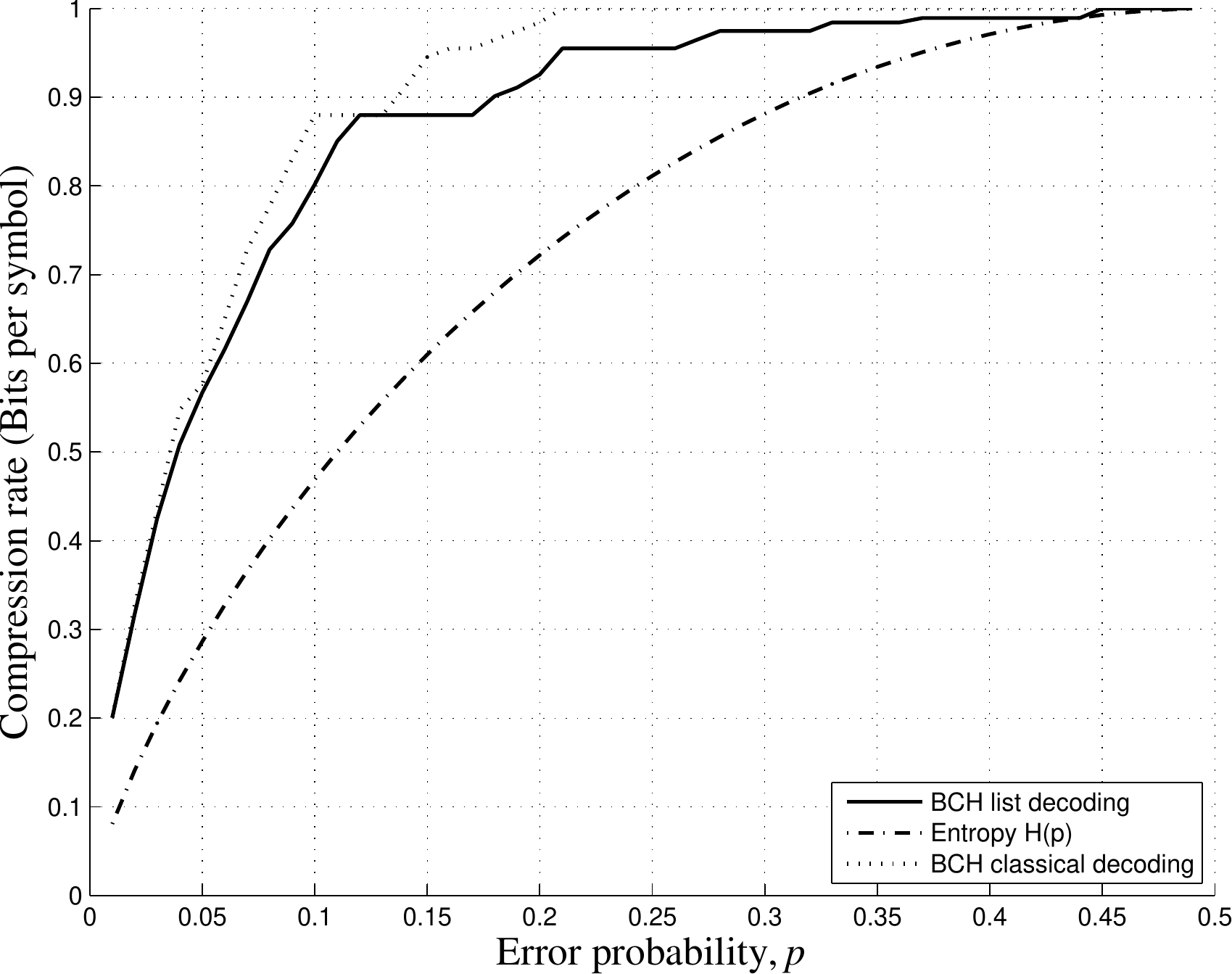}
\caption{Compression rate achievable with list decoding and unique decoding of BCH codes under binary symmetric correlation specified by the crossover probability~$p$ for a desired error rate of $\epsilon < 10^{-3}$.}
\label{fig:SCSI-BCH}
\end{figure}

\ruimte
In Fig.~\ref{fig:SCSI-BCH}, we present the compression rates achievable with BCH codes for binary sources. Here we have used BCH codes of block length $n=1023$ and assumed $\epsilon=10^{-3}$ as the desired error rate. Clearly the SCSI scheme using list decoding of BCH codes outperforms the scheme that relies on classical unique decoding. It is known that for binary sources no compression can be achieved with classical unique decoding if $p \geq 0.25$. It is worth noting that with BCH codes no compression can be achieved with unique decoding if $p>0.20$. For example, when $p=0.21$, the value of $T_\epsilon$ turns out to be $256$. With classical unique decoding to achieve a decoding error rate of $<10^{-3}$ we need a BCH code having a designed  error correcting capability of $t \geq 256$. However, among all the BCH codes of block length $n=1023$, the $(1023, 11)$ BCH code has the maximum designed error correcting capability of $t=255$. Thus we can not achieve any compression using unique decoding of BCH codes when $p>0.20$.

\section{Conclusions}
\label{sec:con}
In SCSI it is customary to model the correlation between the source and side information via a virtual channel. In this paper, we have recognized two important aspects of this virtual channel. Firstly, the compressed symbols are not transmitted over a real noisy channel in the setting of SCSI. We exploit this advantage by transmitting a few additional symbols to accomplish selection of the correct data from the list obtained by the list decoder. These additional symbols are provided by a CRC code. Secondly, we observe that no modulation and demodulation processes are involved in SCSI and thus computationally complex soft-decision decoding is not justified over its hard-decision counterparts. 

\ruimte
We show that our list decoding based source coding has a compression rate that is significantly lower than a classical unique decoding based source coder. Moreover, the proposed approach has the advantage that given the conditional entropy and a class of list decodable code, it allows for an algebraic approach to choosing the best channel code (without resorting to simulation) that achieves the desired error rate. Our future works aim at extending this list-decoding based SCSI scheme to the problem of compression of multiple correlated sources along the line of the approach in~\cite{Margreta-Cao-ICICS09,Margreta-Cao-ISITA,CaoThesis}.
 
\bibliographystyle{plain}
\bibliography{IEEEabrv,MA-MK}

\end{document}